# The *B–X* and *C–X* Band Systems of CuCl Revisited: A Laser-induced Fluorescence Study in 465–490 nm

Li Wang, Junfeng Zhen, Qun Zhang*, Yang Chen*

*Hefei National Laboratory for Physical Sciences at the Microscale and Department of Chemical Physics,*

*University of Science and Technology of China, Hefei, Anhui 230026, People's Republic of China*

*Authors to whom correspondence should be addressed.　E-mail: qunzh@ustc.edu.cn and yangchen@ustc.edu.cn

## ABSTRACT

We have reinvestigated the $B-X$ and $C-X$ band systems of CuCl by recording the laser-induced fluorescence excitation spectra in 20400–21800 cm$^{-1}$. The rotational analyses in Hund's case (a) revealed unambiguously a singlet-to-singlet transition nature. The measured lifetimes of a few microseconds seem too long for singlets and too short for triplets, which we think is actually in favor of a picture of singlet ($^1\Pi$ and $^1\Sigma^+$)-triplet ($^3\Pi_{0,1,2}$) mixed states in the *B* and *C* band systems of CuCl. The two excited states we observed in our spectra may be the singlets that have been strongly "contaminated" by their triplet neighbors.

The visible spectrum of CuCl was first observed in 1927 by Ritschl [1] and in 1938 by Bloomenthal [2] who classified the observed bands into five systems, namely, $A$(5150–5500 Å), $B$(4620–5110 Å), $C$(4660–5170 Å), $D$(4140–4530 Å), and $E$(3990–4580 Å), all of them having a common lower state. A new band system, $F$(3790–4180 Å), was reported later by Rao *et al.* [3] The $B$ and $C$ band systems of $^{63}Cu^{35}Cl$ lying in the region 460–520 nm were first rotationally analyzed by Lagerqvist and Lazarava-Girsamoff [4] and Rao *et al.* [5] who assigned them to the $B^1\Pi - X^1\Sigma^+$ and $C^1\Sigma^+ - X^1\Sigma^+$ transitions. In 1987 Delaval *et al.* [6] argued that the two excited singlets may be reassigned to $B^3\Pi_1$ and $C^3\Pi_0$, components of the 3$d$-hole $^3\Pi$ state. This argument was made based mainly on the measured radiative lifetimes (~3 μs for both states [6]) and supported, to a certain extent, by their lifetime calculations [7] using a model including spin-orbit interaction mixings within the $Cu^+(3d^94s)Cl^-(3s^23p^6)$ structure. Following this spirit as well as taking into account *ab initio* calculations by Sousa *et al.*[8], Parekunnel *et al.* [9] attributed their Fourier transform emission spectra in the $B - X$ and $C - X$ band systems to triplet-singlet transitions and further relocated the $B^1\Pi$ and $C^1\Sigma^+$ states [4,5], or the $B^3\Pi_1$ and $C^3\Pi_0$ states [6-8], to the $b^3\Pi_1$ and $b^3\Pi_0$ states, respectively.

It is known that in copper compounds large spin-orbit effects allow otherwise spin-forbidden triplet-singlet transitions to occur, showing smaller spectral intensities than allowed singlet-singlet ones [10,11]. This indeed lends support to the triplet-singlet assignments of the CuCl emission spectra [9]. Nevertheless, the presence of triplet-singlet transitions, forbidden by Hund's case (a) selection rules, suggests that there is considerable mixing of the triplet and singlet states. This implies that it may not be convincing to relocate the singlets ($B^1\Pi$ and $C^1\Sigma^+$ [4,5]) to the triplets (either $B^3\Pi_1$ and $C^3\Pi_0$ [6-8] or $b^3\Pi_1$ and $b^3\Pi_0$ [9]).

As demonstrated in our recent work [12] on the well-known Na$_2$ $b^3\Pi_{\Omega u} \sim A^1\Sigma_u^+ - X^1\Sigma_g^+$ inter-combination band system, spectroscopic detection can be readily "switched" between the spin-allowed singlet-singlet $A^1\Sigma_u^+ - X^1\Sigma_g^+$ transition and the spin-forbidden triplet-singlet $b^3\Pi_{\Omega u} - X^1\Sigma_g^+$ transition. We anticipate that for CuCl the $b^3\Pi_{0,1} - X^1\Sigma^+$ transitions observed by Parekunnel et al. [9], similar to the case in Na$_2$ [12], may concur with the $B^1\Pi - X^1\Sigma^+$ and $C^1\Sigma^+ - X^1\Sigma^+$ transitions observed by Lagerqvist and Lazarava-Girsamoff [4] and Rao et al. [5]. We therefore suspect that the substitution of the two singlets ($B^1\Pi$ and $C^1\Sigma^+$) with the two triplets ($b^3\Pi_1$ and $b^3\Pi_0$) in CuCl [9] could be invalid.

With this suspicion in mind we revisited, by means of laser-induced fluorescence (LIF) excitation spectroscopy, the spectral region (465–490 nm) in which the *B* and *C* band systems are expected to lie. Our experiment was conducted in a LIF apparatus which has been described elsewhere [13]. The CuCl molecules were produced by the reaction of HCl gas (2% seeded in argon) with the copper atoms sputtered from a pair of pure copper pin electrodes under pulsed DC discharge and supersonic expansion conditions. Figure 1 shows the survey LIF excitation spectrum recorded in 20400–21480 cm$^{-1}$. Based on previous reports [4,5,9] we can readily assign the spectral features to the 0–0, 1–0, and 2–0 bands of the $B - X$ and $C - X$ systems, as labeled by ticks in Fig.1.

The rotationally resolved spectra together with the simulations using the PGOPHER software [14] for the two 0–0 bands are shown in Figs. 2(a) and (b), both exhibiting good agreement between the observed and simulated spectra. They revealed a rotational structure typical of transitions $\Pi - \Sigma$ [showing *P*, *Q*, and *R* branches in Fig. 2(a)] and $\Sigma - \Sigma$ [showing only *P* and *R* branches in Fig. 2(b)]. Considering that the common lower state is the singlet $X^1\Sigma^+$ state [4,5,9] and our simulations were performed in Hund's case (a), we can unambiguously assign the two 0–0 bands to the $B^1\Pi - X^1\Sigma^+$ and $C^1\Sigma^+ - X^1\Sigma^+$ transitions.

Limited by paper length, among the remaining bands we show only the $B^1\Pi - X^1\Sigma^+$ 1–0 band and the $C^1\Sigma^+ - X^1\Sigma^+$ 2–0 band in Figs. 3(a) and (b), respectively. Both spectra clearly show features arising from the four isotopmers of CuCl ($^{63}$Cu$^{35}$Cl, $^{65}$Cu$^{35}$Cl, $^{63}$Cu$^{37}$Cl, and $^{65}$Cu$^{37}$Cl with a natural abundance ratio of ~7:3:2.25:1). Because of the severe spectral blending only the R heads for $^{65}$Cu$^{35}$Cl, $^{63}$Cu$^{37}$Cl, and $^{65}$Cu$^{37}$Cl were marked by ticks in Figs. 3(a) and (b), from which we can see again a nice match between the observed and simulated spectra. This in turn confirms that the spectral carrier is indeed CuCl. More importantly, all the observed isotope shift values were found to match nicely with the calculated ones, a feat that was achieved in none of the previous literature [4,5,9]; this further made us confident of our spectral assignments. In addition, the resultant spectroscopic constants ($T_e = 20\,484.6\,\text{cm}^{-1}$ and $B_e = 0.1699\,\text{cm}^{-1}$ for $B^1\Pi$; $T_e = 20\,631.7\,\text{cm}^{-1}$ and $B_e = 0.1711\,\text{cm}^{-1}$ for $C^1\Sigma^+$) were also found to agree well with those reported in Ref. 5. All the spectroscopic data we obtained are not listed here due to the space limit (see supplementary documents).

It is noteworthy that the measured radiative lifetimes for the $B^1\Pi\,(\upsilon'=0)$ and $C^1\Sigma^+\,(\upsilon'=0)$ states (4.670 and 4.667 μs, respectively) are of the same order of magnitude as those for the $B^3\Pi_1$ and $C^3\Pi_0$ states (3.3 and 3.2 μs, respectively [6]). It was hoped that the determination of the radiative lifetimes of CuCl would make it possible to distinguish, in principle, between the triplet and singlet excited states, the singlets being expected to exhibit much shorter lifetimes than the triplet ones [6,11]. Unfortunately, the lifetimes of a few microseconds appear too long for singlets while too short for triplets. Nevertheless, we think this is actually in favor of a mixed-states picture [15] of the coexistence between the singlets ($B^1\Pi$ and $C^1\Sigma^+$) and the triplets ($^3\Pi_1$ and $^3\Pi_0$, both labeled *b* by Parekunnel *et al.* [9]; perhaps also including the "missing" $^3\Pi_2$ component). Therefore, the two singlets we detected are supposed to be strongly perturbated *via* strong spin-orbit couplings by their triplet neighbors. It is the singlet-triplet mixings that make longer than expected the lifetimes of our observed singlets and allow otherwise spin-forbidden triplet-singlet transitions to occur [9].

To summarize, we have reinvestigated the $B-X$ and $C-X$ band systems of CuCl by recording the LIF excitation spectra in 20400–21800 cm$^{-1}$, revealing an unambiguous singlet-to-singlet transition nature. Indeed, the measured lifetimes of a few microseconds seem too long for singlets and too short for triplets, which we think actually supports a picture of singlet ($^1\Pi$ and $^1\Sigma^+$)-triplet ($^3\Pi_{0,1,2}$) mixed states in the *B* and *C* band systems of CuCl.

The authors acknowledge the support from the National Natural Science Foundation of China (Grant Nos. 20673107 and 20873133), the National Key Basic Research Special Foundation of China (Grant Nos. G2007CB815203 and 2010CB923302), and the Chinese Academy of Sciences (KJCX2-YW-N24).

# Figure Captions

FIG. 1.  The survey LIF excitation spectrum of the $B-X$ and $C-X$ band systems of CuCl in 20 400 – 21 800 cm$^{-1}$.

FIG. 2.  Rotationally resolved LIF excitation spectrum of the (0, 0) bands for (a) $B^1\Pi - X^1\Sigma^+$ and (b) $C^1\Sigma^+ - X^1\Sigma^+$ transitions of CuCl.  The observed rotational lines for $^{63}$Cu$^{35}$Cl are indicated by ticks with corresponding $J$ numbers.

FIG. 3.  Rotationally resolved LIF excitation spectrum of (a) the $B^1\Pi - X^1\Sigma^+$ (1, 0) band and (b) the $C^1\Sigma^+ - X^1\Sigma^+$ (2, 0) band of CuCl.  For the isotopomers $^{65}$Cu$^{35}$Cl, $^{63}$Cu$^{37}$Cl, and $^{65}$Cu$^{37}$Cl, only the $R$ heads are indicated by ticks due to severe spectral blending.  The observed rotational lines for $^{63}$Cu$^{35}$Cl are indicated by ticks with corresponding $J$ numbers.

**Fig. 1    (Wang *et al.*)**

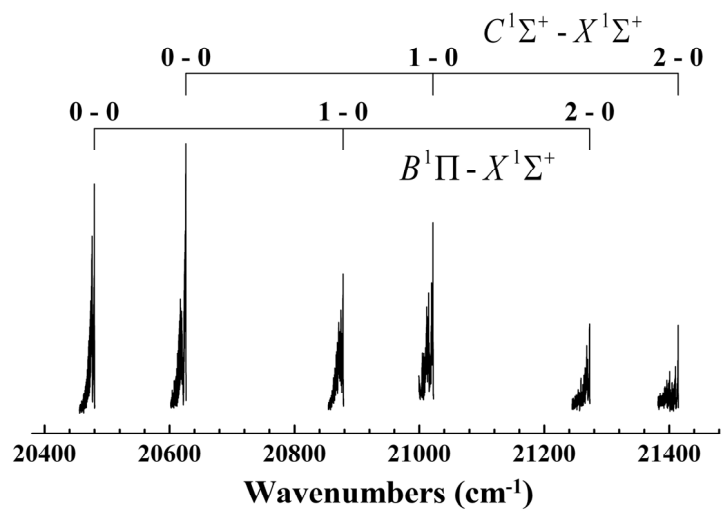

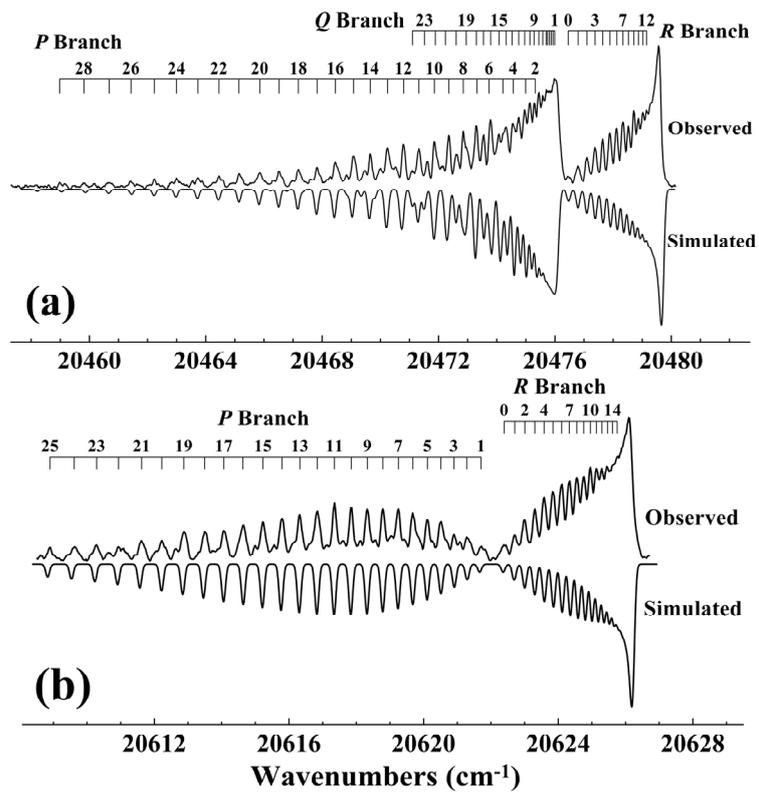

Fig. 3 (Wang *et al.*)

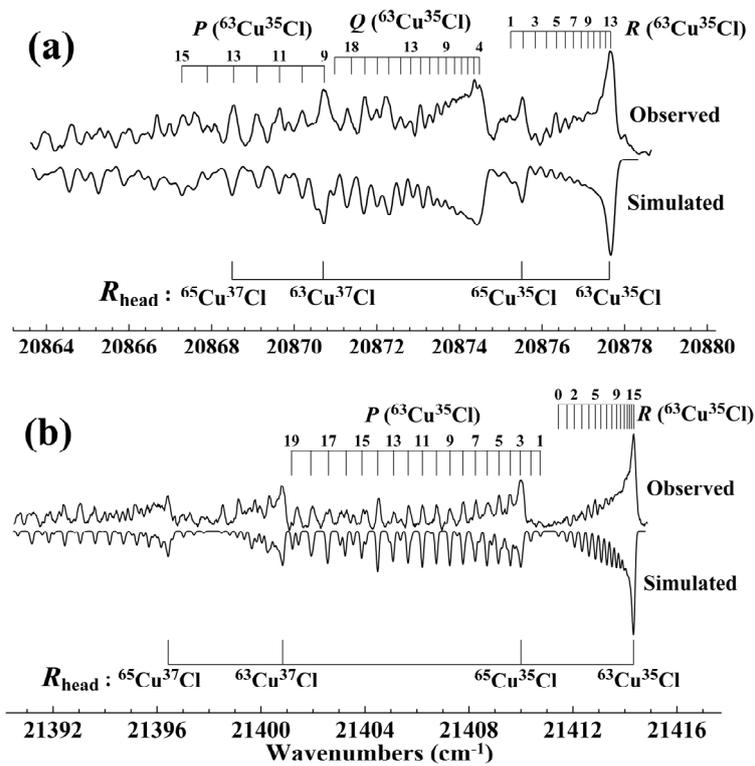